# Title page

UltraEar: a multicentric, large-scale database combining ultra-high-resolution computed tomography and clinical data for ear diseases

**Authors:**

[1] Ruowei Tang, Ph.D., E-mail: tangrw@ccmu.edu.cn

[1] Pengfei Zhao[*], Ph.D., E-mail: zhaopengf05@163.com, ORCID: 0000-0002-9210-6544

[2] Xiaoguang Li, Ph.D., E-mail: lxg@bjut.edu.cn

[1] Ning Xu, M.D., E-mail: xn18548231899@163.com

[1] Yue Cheng, M.D., E-mail: cy13855971620@163.com

[3] Mengshi Zhang, Ph.D., E-mail: zhangmengshi17@163.com

[4] Zhixiang Wang, Ph.D., E-mail: zhwang93@163.com

[1] Zhengyu Zhang, Ph.D., E-mail: zhangzy203@163.com

[1] Hongxia Yin, M.D, E-mail: 282496774@qq.com

[1] Heyu Ding, Ph.D., E-mail: dingheyu1987@163.com

[5] Shusheng Gong, Ph.D., E-mail: gongss1962@163.com

[5] Yuhe Liu, Ph.D., E-mail: liuyuhefeng@163.com

[1] Zhenchang Wang, M.D., Ph.D., E-mail: cjr.wzhch@vip.163.com, ORCID: 0000-0001-8190-6469

**Affiliations:**

[1] Department of Radiology, Beijing Friendship Hospital, Capital Medical University, Beijing, China, 100050

[2] Beijing Key Laboratory of Computational Intelligence and Intelligent System, Beijing University of Technology, Beijing, 100124, China

[3] Yanjing Medical College, Capital Medical University, Beijing 101300, China

[4] Department of Ultrasound, Beijing Friendship Hospital, Capital Medical University, Beijing 100050, China

[5] Department of Otolaryngology and Head & Neck, Beijing Friendship Hospital, Capital Medical University, Beijing, China, 100050

# UltraEar: a multicentric, large-scale database combining ultra-high-resolution computed tomography and clinical data for ear diseases

**Abstract**

Ear diseases affect billions of people worldwide, leading to substantial health and socioeconomic burdens. Computed tomography plays a pivotal role in accurate diagnosis, treatment planning, and outcome evaluation. The objective of this study is to present the establishment and design of UltraEar Database, a large-scale, multicentric repository of isotropic 0.1 mm ultra-high-resolution computed tomography (U-HRCT) images and associated clinical data dedicated to ear diseases. UltraEar is an ongoing retrospective and prospective study, recruiting patients from 11 tertiary hospitals between October 2020 and October 2035. The database integrates U-HRCT images, structured CT reports, and comprehensive clinical information, including demographics, audiometric profiles, surgical records, and pathological findings. A broad spectrum of otologic disorders is covered, such as otitis media, cholesteatoma, ossicular chain malformation, temporal bone fracture, inner ear malformation, cochlear aperture stenosis, enlarged vestibular aqueduct, and sigmoid sinus bony deficiency. Standardized preprocessing pipelines have been developed for geometric calibration, expert annotation, and multi-structure segmentation. All personal identifiers in DICOM headers and metadata are removed or anonymized to ensure compliance with data privacy regulations. Data collection and curation are coordinated through monthly expert panel meetings, with secure storage on an offline cloud system to safeguard integrity. As the first comprehensive isotropic 0.1 mm U-HRCT database targeting ear diseases, UltraEar database provides an unprecedented ultra-high-resolution reference atlas with both technical fidelity and clinical relevance. This resource has significant potential to advance radiological research, enable development and validation of artificial intelligence algorithms, serve as an educational tool for training in otologic imaging, and support multi-institutional collaborative studies. UltraEar Database will be continuously updated and expanded, ensuring long-term accessibility and usability for the global otologic research community. Registry: ClinicalTrials.gov, TRN: NCT05533840, Registration date: September 5[th], 2022.

**Abbreviations**

| | |
|---|---|
| CT | Computed tomography |
| MSCT | Multislice CT |
| PCD-CT | Phonton-counting detector CT |
| U-HRCT | Ultra-high-resolution CT |
| AI | Artificial intelligence |

| SOP | Standard observation plane |
| DICOM | Digital Imaging and Communications in Medicine |
| CBCT | Cone-beam CT |

# BACKGROUND

Ear diseases, with the symptoms such as hearing loss, tinnitus, vertigo and facial paralysis, pose medical burden worldwide. As GBD 2021 indicates, prevalence of hearing loss increases to ~20.3% (1.57 billion) of world population in 2019, and a projected prevalence of 2.45 billion in 2050. And tinnitus affects 11.2% of the United States population, with 28.3% of those affected having a symptom duration of more than 15 years[1].

The ear contains delicate structures with key anatomical components located deep within the temporal bone. Computed tomography (CT) examination is a pivotal non-invasive method for visualizing the causative factors of otological diseases. However, due to the intricate nature of ear structures, the dimensions of the detector constrain the spatial resolution of a multislice CT (MSCT) scanner to ~0.625 mm. This limitation impedes the clear display of crucial substructures and accurate diagnosis of occult lesions, since the smallest structures of the middle ear, i.e., the stapes footplate, could be as thin as 0.18 mm[2]. And 45.4% cases of otosclerosis, manifesting as a focal low-attenuated lesion of the fissula ante fenestram, might be missed on MSCT[3]. Furthermore, the presence of numerous anatomical features and variations within the temporal bone such as the cochlear fissure, can result in erroneous interpretations of pathological changes. Another factor contributing to misdiagnosis of ear diseases is the lack of experience interpreting otologic imaging, particularly among radiologists who do not specialize in head and neck imaging.

Technological advancements have made significant increases in spatial resolution possible. Phonton-counting detector CT (PCD-CT) offers efficient ultra-high-resolution imaging with improved iodine contrast-to-noise characteristics and multi-energy material decomposition, with a maximal resolution of 0.15 mm. Preliminary research has been reported to image the anatomic structures and some lesions of the ear[4, 5]. However, most studies on PCD-CT focus on technical validation rather than on its diagnostic performance. And the high equipment cost, increased data volume, and compatibility requirements for image processing pose practical challenges for PCD-CT [6-8]. Meanwhile, a newly developed device, named ultra-high-resolution CT (U-HRCT), is used to depict anatomical features and identify occult ear diseases in healthy volunteers and patients[9]. Incorporating a small-focus X-ray generator and a flat panel detector, U-HRCT realized an isotropic resolution of 0.1 mm.

At present, publicly available databases for ear are limited in size and scope. Existing resources such as OpenEar[10], Visible Human Project[11], and Visible Ear[12], are small in sample size or on tissue sectioning. Database incorporating both imaging and clinical data for patients with ear diseases has not been reported yet. Therefore, we established a multicenter, large-scale repository (UltraEar database) of isotropic 0.1 mm ultra-high-resolution CT images and clinical data for a broad spectrum of ear diseases. The aims are to: (1) evaluate the diagnostic efficacy of U-HRCT for ear diseases, (2) establish new diagnostic standards for lesion detection, symptom correlation, and postoperative follow-up, and (3) serve as a high-resolution reference resource to support clinical research, artificial intelligence (AI) development, and education in otologic imaging. This paper gives a comprehensive overview of UltraEar

Database.

## CONSTRUCTION AND CONTENT
### Study design
This is an ongoing multicentric, retrospective and prospective study including patients with ear diseases at Beijing Friendship Hospital, Capital Medical University and other 10 tertiary referral centers across China. We intend to enroll 30000 patients from October 2020 to October 2035. To date, more than 8000 patients have been enrolled in UltraEar database (Figure 1).

### Study population
Patients are included with the following inclusion criterion: admitted to the Otolaryngology and Head & Neck Department for otologic symptoms, such as otorrhea, hearing loss, tinnitus, otalgia, ear fullness, otogenic vertigo, facial paralysis and postoperative re-examination. The exclusion criteria are as follows: (1) patients who are vulnerable to radiation exposure, such as pregnant and lactating women; (2) patients who are unable to cooperate with U-HRCT or clinical examinations; (3) patients whose images have motion artifacts that interfere with image interpretation. A spectrum of ear diseases is included in UltraEar Database, including otitis media, cholesteatoma, ossicular chain malformation, temporal bone fracture, inner ear malformation, cochlear aperture stenosis, enlarged vestibular aqueduct, and sigmoid sinus bony deficiency (Figure 2).

### Data collection
### U-HRCT protocol
The temporal bones were scanned unilaterally with a U-HRCT scanner (Ultra3D, LargeV, Beijing) at 100–110 kVp, 120–180 mAs, and both temporal bones of each patient were scanned. The focal spot size of the X-ray source was 0.25 mm (IEC 60336) and the detector pixel size was 0.0748 mm × 0.0748 mm. The device incorporated a 35-mm collimation, a field of view of 65 mm, and a voxel size of 0.1 mm × 0.1 mm × 0.1 mm. The U-HRCT scanning generated isotropic voxels, enabling visualization of the temporal bone structures from any desired direction. The exposure dose ranged from 156–295 μGy $m^2$ for each scan.

### CT report
A system for generating structured CT reporting is used, and the CT report contains the following anatomical structures: external auditory canal, tympanic membrane, tympanic cavity wall, the ossicular chain, facial nerve canal, cochlea, vestibule, semicircular canals, internal auditory canal, oval window, round window, jugular bulb fossa, sigmoid sinus, cochlear aqueduct, vestibular aqueduct and middle cranial fossa. Free-text and structured CT reports containing structures and common descriptions and diagnoses are listed in Table 1 and Supplemental Table, respectively.

A two-step protocol has been established for CT report drafting: an initial draft by a junior neuroradiologist, followed by a second review by a senior neuroradiologist with >5 years of experience. All radiologists who participate in the draft of CT report are trained by at least one senior neuroradiologist with 2-hour training every two weeks for 6 months.

**Clinical data**

Clinical data are retrieved through the off-line clinical database, which are categorized as follows: (1) demographic data such as age, sex, ethnicity; (2) chief complaint, relevant symptoms and family history; (3) for patients with hearing loss/otorrhea/ear pain/ear fullness, auditory testing such as pure tone audiometry, tympanometry, auditory brainstem response and speech audiometry; (4) for patients with tinnitus, additional questionnaires such as tinnitus handicap inventory, pitch matching testing and visual analog scale; (5) for patients with otogenic vertigo, additional testing such as videonystagmography, vestibular evoked myogenic potentials and vertigo symptom scale; (6) for patients with peripheral facial paralysis/facial spasm, additional testing such as House-Brackmann grading scale; (7) surgical data, including surgical procedure, surgical finding and surgical time; and (8) pathological findings.

**Data processing**
**U-HRCT images**
1. Geometric calibration and standard observation plane (SOP)

Due to variations in subject posture and CT scanner settings, U-HRCT images require geometric alignment through multiplanar reconstruction to ensure bilateral anatomical symmetry[9]. An improved geometric calibration algorithm based on segmentation of the lateral semicircular canal is developed, where a plane is fitted using binary image refinement and skeletonization. The normal vector of this plane defines the Z-axis. The skeleton extraction algorithm refines connected regions to single-pixel width for robust feature extraction and topological representation. Two endpoints of the lateral semicircular canal are then identified, and the straight line connecting them is used to estimate the rotation angle about the Z-axis. The angle between this line and the mid-sagittal plane is estimated using the average angle derived from a set of bilateral temporal bone CT images (Figure 3).

SOPs are essential for temporal bone CT diagnosis, particularly for evaluating the delicate ossicular chain. Eleven SOPs are defined for the ossicular chain. Based on segmentation of the ossicles, an SOP-Graph model is implemented to represent the spatial relationships between different SOPs and to encode radiologists' search strategies [13].

2. Image annotation

A dedicated annotation team comprising neuroradiologtists is established, which contains at least 2 senior radiologists. All annotators have a solid foundation in medical imaging, with specific expertise in otologic anatomy and U-HRCT image interpretation. The annotators independently label the malleus, incus, stapes, cochlea, semicircular canals, and vestibule on U-HRCT images using Mimics (version 19.0, Materialise NV, Leuven, Belgium), while being blinded to clinical information. The annotation workflow is as follows: (1) import Digital Imaging and Communications in Medicine (DICOM) images into Mimics; (2) manually delineate the contours of each anatomical structure and exclude adjacent soft tissues and artifacts; (3) perform manual morphological refinement of region-of-interest boundaries; and (4) export the finalized annotations in .zip format (Figure 4).

3. Structure segmentation

A deep learning framework integrating multi-view fusion with active contour constraints was developed for automatic segmentation in U-HRCT. The model employs TransUNet, a hybrid encoder-decoder architecture that combines Transformer modules with conventional convolutional neural networks. To leverage complementary information from multiplanar reconstructions, a fusion strategy synchronizing coronal, sagittal, and axial views is implemented. Specifically, for the delicate stapes, an active contour loss function based on the Euler curve energy formulation is designed. This function incorporates curvature and contour length constraints to enforce topological integrity, operating synergistically with standard Dice and cross-entropy losses during end-to-end training.

**CT reporting**

The CT report texts are acquired through the Hospital Information System and the Radiology Information System and then underwent preprocessing. Patient privacy information is de-identified. Expressions such as high-density shadow are normalized to the standard term high-density attenuation. Chinese word segmentation tools are used for tokenization and stop-word filtering. A three-level label hierarchy is constructed based on the International Classification of Diseases, 10th Revision and relevant clinical guidelines. The first level comprises major disease categories such as external ear and middle ear conditions; the second level include subcategories such as external ear malformation and middle ear inflammation; and the third level encompasses specific diseases and imaging findings, including external auditory canal atresia and chronic otitis media. In addition, the system incorporates imaging feature labels such as high-density attenuation and anatomical structure labels such as the malleus and cochlea. The Qwen-1.8B large language model is used to directly extract labels. Its architecture involves feeding tokenized CT report texts into the model to generate labels, leveraging contextual semantics to output structured results. The multimodal label table construction process involves extracting imaging labels from CT reports, mapping clinical diagnoses to standardized clinical labels, and integrating both to create a unified label table containing both imaging and clinical information.

**Sample size estimation**

The PASS software (Version 15.0.5, NCSS, LLC) is used to sample size calculation. Otosclerosis, a primary focal osteodystrophy of the otic capsule with a focal lesion size less than 1.0 mm, is referenced by statistician as an example for enhancing diagnostic precision. Based on our previous study, we assume that the sensitivity for identifying otosclerosis is 80%. With an allowable error ($\delta$) of 0.075, a type I error ($\alpha$) of 0.05, a type II error ($\beta$) of 0.2 and power of 80%, this calculation indicates that 20 confirmed otosclerosis cases are needed. With an estimated prevalence of 0.3% in the general population, the sample size for the general population is 6667. Assuming a 20% loss to the follow-up rate, the estimated minimum sample size needed for this study is approximately 8334 individuals.

**Statistical Analysis**

Statistical analyses were performed with SPSS 26.0 (IBM Inc., Chicago, Illinois). Quantitative data are presented as mean standard deviation. Differences in anatomic variables between paired ears are calculated with the paired *t*-test for normal distribution, and Wilcoxon rank-test for abnormal distribution. Differences between men and women in terms of quantitative measurement of the anatomic variables are identified using the independent *t*-test or Mann-Whitney *U* test. Statistical differences are considered significant with a two-tailed $P < 0.05$.

**Quality control of the database**

All participating centers are equipped with the same brand and model of U-HRCT devices, with identical CT scanning parameters, as described above. To ensure patient confidentiality and comply with ethical standards, all patient identifiers within the DICOM image headers and associated data, including patient name, ID number, date of birth, and institution-specific information are systematically removed or obscured prior to analysis. The expert panel (comprising ≥5 senior neuroradiologists) holds monthly communication meeting to discuss over the image quality, CT report, and normal/diseased ear structures to ensure accuracy and consistency in diagnosis. Data are collected monthly and securely stored on an offline cloud drive. U-HRCT and clinical data are stored at the institution where the data are collected, and an additional one copy is maintained at Beijing Friendship Hospital, Capital Medical University. Data supervision is conducted by all the principal investigators from all the centers after professional training.

**UTILITY AND DISCUSSION**

UltraEar database presents the first large-scale, comprehensive isotropic 0.1 mm U-HRCT images and the associated clinical data specifically targeting ear diseases. Compared to conventional MSCT with slice thicknesses ranging from 0.5 to 0.625 mm, U-HRCT shows improved spatial resolution, enabling detailed visualization of critical structures and accurate diagnosis of occult lesions. This level of resolution, previously achievable only in micro-CT or cadaveric studies, is now available in vivo, allowing analysis of anatomical variability, early pathology, and surgical planning.

The acquisition protocol was standardized across all scans at the participating centers to ensure data with high homogeneity. All data underwent strict quality control, including multi-phase expert review by radiologists, in image enrollment, CT report and image annotation. This ensures both technical quality and clinical usability, supporting applications in research, surgical simulation, and AI development.

UltraEar database has the potential to serve as a high-resolution reference atlas for otologic imaging, supporting both diagnostic, research and educational purposes. It enhances the ability to delineate subtle osseous changes that may be involved in otosclerosis[8, 14], congenital ear malformation, Ménière's disease[15], and facial nerve dehiscence[16]. UltraEar database enables radiologists to refine diagnostic criteria, facilitating more precise diagnoses and informed treatment plans. The acquisition of measurements pertaining to structures such as the vestibular aqueduct, cochlear nerve canal, or stapes footplate can be achieved with a degree of precision that surpasses prior limitations.

To our knowledge, no existing public temporal bone database offers both isotropic 0.1-mm resolution CT images and clinical data. For example, OpenEar datasets, incorporating cone-beam CT (CBCT) and micro-slicing to achieve the effect of temporal bone specimen[10]. Gerber et al. established a multiscale imaging and modelling dataset of the human inner ear based on 52 temporal bones scanned with CBCT and micro-CT with 30 and 50 image volumes, respectively[17]. Ear Imagery Database using otoscopic images to classify normal, myringosclerosis, earwax plug, and chronic otitis media[18], while use a large database of 10544 otoendoscopic images to identify ear diseases[19]. The human bony labyrinth dataset comprising 23 specimens with respective 0.2-mm CT and 0.06-mm μCT images[20]. National Temporal Bone, Hearing, and Balance Pathology Resource Registry[21], Visible Human Project[11] and Visible Ear[12] primarily focus on otopathologic research. Genetic resources like the Deafness Variation Database provides genetic, genomic, and clinical data for deafness with no imaging data. The datasets above are either on pathological results, or with limited radiological applicability. They are often designed for anatomical modeling or surgical training, and lack standard clinical annotations necessary for the development of AI models. In contrast, UltraEar database is expected to accelerate algorithm development, benchmarking, and clinical translation in ear diseasess.

PCD-CT is suitable for imaging ear diseases such as ossicular malformation[4, 5], superior semicircular canal dehiscence[22], inflammatory diseases and postoperative evaluation[23]. However, the majority of present studies focused on image quality metrics rather than on image interpretation[24, 25], which makes PCD-CT remain in the stage of technical validation. Certain anatomical landmarks may appear differently compared to conventional energy-integrating detector CT, requiring re-training of radiologists and re-visiting of anatomic features. U-HRCT with 0.1 mm isotropic resolution shows its advantage in standardized acquisition and interpretation protocols, making it more readily applicable to clinical diagnosis and research[6, 7]. Therefore, the establishment of UltraEar database can potentially provide a usable diagnostic system for PCD-CT.

In addtion, UltraEar database provides a strong foundation for advancing AI in otologic imaging. With segmentation and anatomical labeling, future AI models using U-HRCT images may detect occult lesion such as early otosclerosis, ossicular malformation, and labyrinthitis ossificans. Based on UltraEar database, several studies on stapes fixation[26], labyrinth segmentation[27] and morphological changes in Ménière's disease[15] have been reported. Ultimately, continuous clinical validation and feedback will ensure these AI tools are robust and generalizable across diverse patient populations and clinical settings. Thus, the synergy between U-HRCT and AI heralds a new era in otologic diagnostics and therapeutics, with UltraEar database serving as a critical resource.

UltraEar database includes free-text and structured CT report. With clear visualization of anatomical subregions, radiologists can adopt structured templates that include quantitative measurements of clinically relevant landmarks. The Audiological and Genetic Database is a public, de-identified research database for pediatric hearing research, containing 15059 radiology reports from 37273 patients[28]. Based on this database, Masino et al.[29] developed a machine learning approach to label the reports with ear abnormalities. With large-scale text data (CT report and clinical data), UltraEar database also has the potential to develop toward text extraction and intelligent report generation in future studies.

Despite the strengths, this study has several limitations. First, long-term follow-up is not sufficient for the entire cohort, which may impede its clinical application. Future version of the database should aim more on patients who undergo surgical procedure, which could provide better follow-up result. Second, although annotations are reviewed by experienced radiologists, inter-observer variability in identifying anatomical landmarks remains a challenge that warrants further investigation.

**Ethics and dissemination**

This study is registered at ClinicalTrials.gov (NCT05533840) on September 5$^{th}$, 2022, and has been approved by local ethnic committee of Beijing Friendship Hospital, Capital Medical University (approval numbers: 2018-P2-210-02, 2020-P2-061-02, 2022-P2-055-01 and 2024-P2-061-02). Written informed consent is obtained from each participant, or legal guardians. Modifications to the protocol will be communicated to all involved parties and approved by the ethics committee to ensure transparency and compliance. Privacy and confidentiality of study participants will be preserved in all reports and publications. Authorship eligibility is presented following the International Committee of Medical Journal Editors' four criteria.

**CONCLUSIONS**

In conclusion, UltraEar Database is a multicentric, large-scale joint imaging-clinical database, aiming to provide a high-resolution atlas for advancing radiological research, enable development and validation of artificial intelligence algorithms. UltraEar Database will be continuously updated and expanded, ensuring long-term accessibility and usability for the global otologic research community.

# Figure

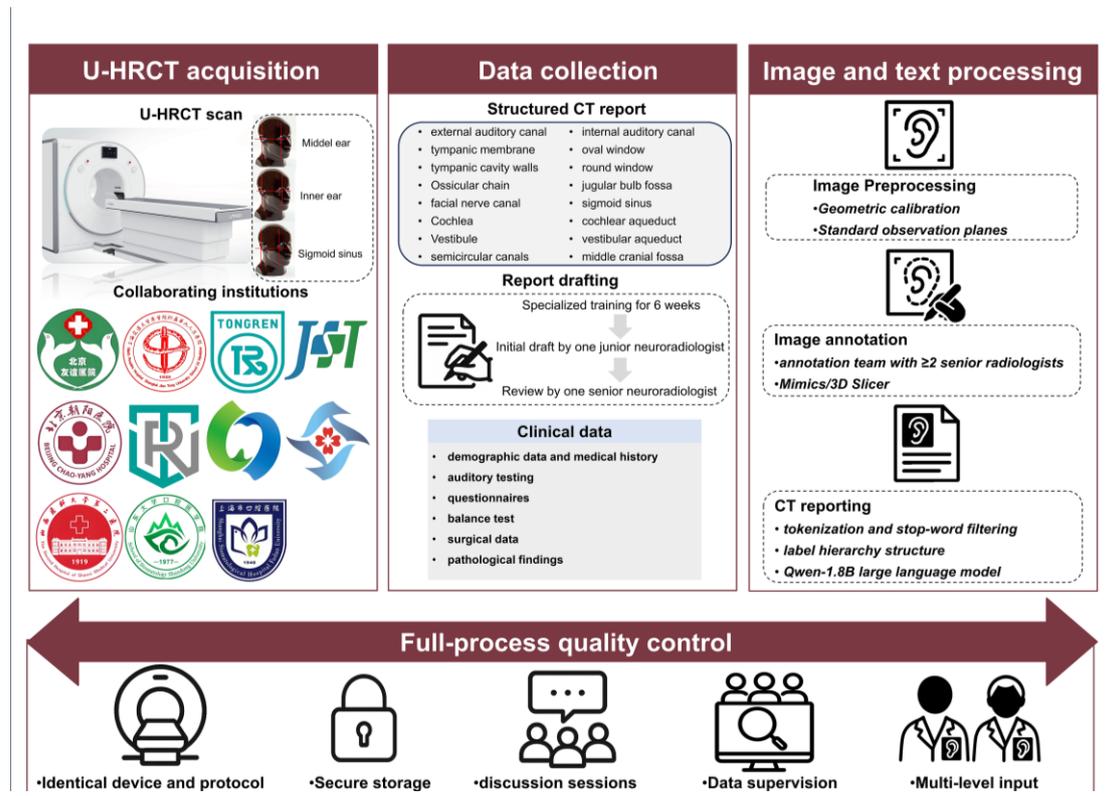

Figure 1. The flowchart of UltraEar database establishment and full-process quality control

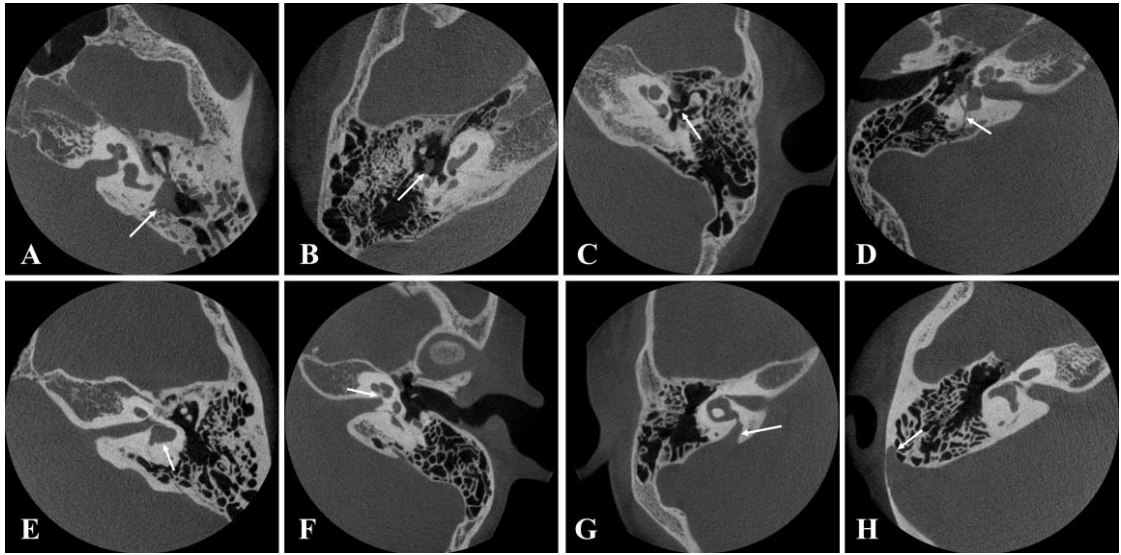

Figure 2. A spectrum of ear diseases is included in UltraEar Database, including otitis media (A), cholesteatoma (B), ossicular chain malformation (C), temporal bone fracture (D), inner ear malformation (E), cochlear aperture stenosis (F), enlarged vestibular aqueduct (G), and sigmoid sinus bony deficiency (H). Abnormal findings are illustrated by arrows

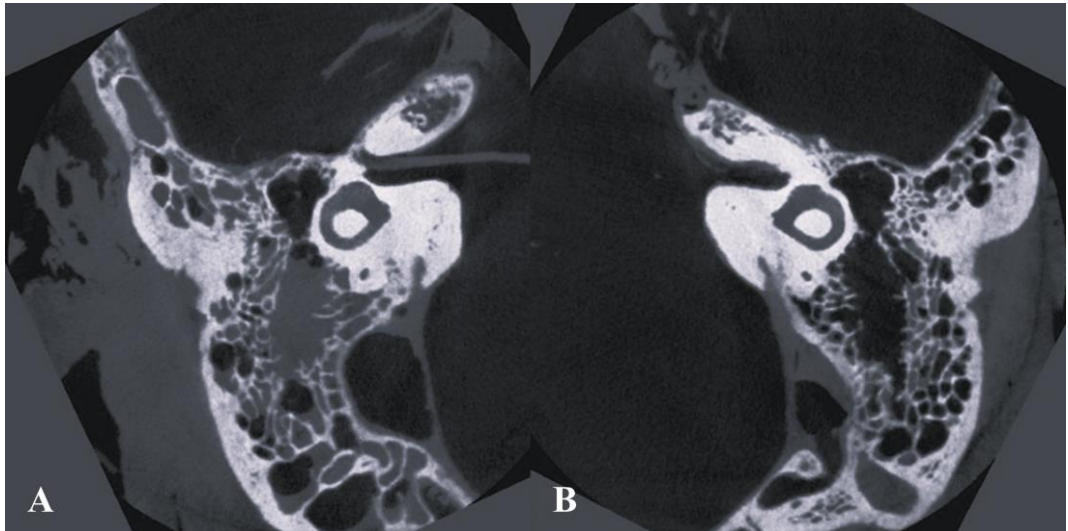

Figure 3. The results of geometric calibrated U-HRCT images fitting the lateral semicircular canal

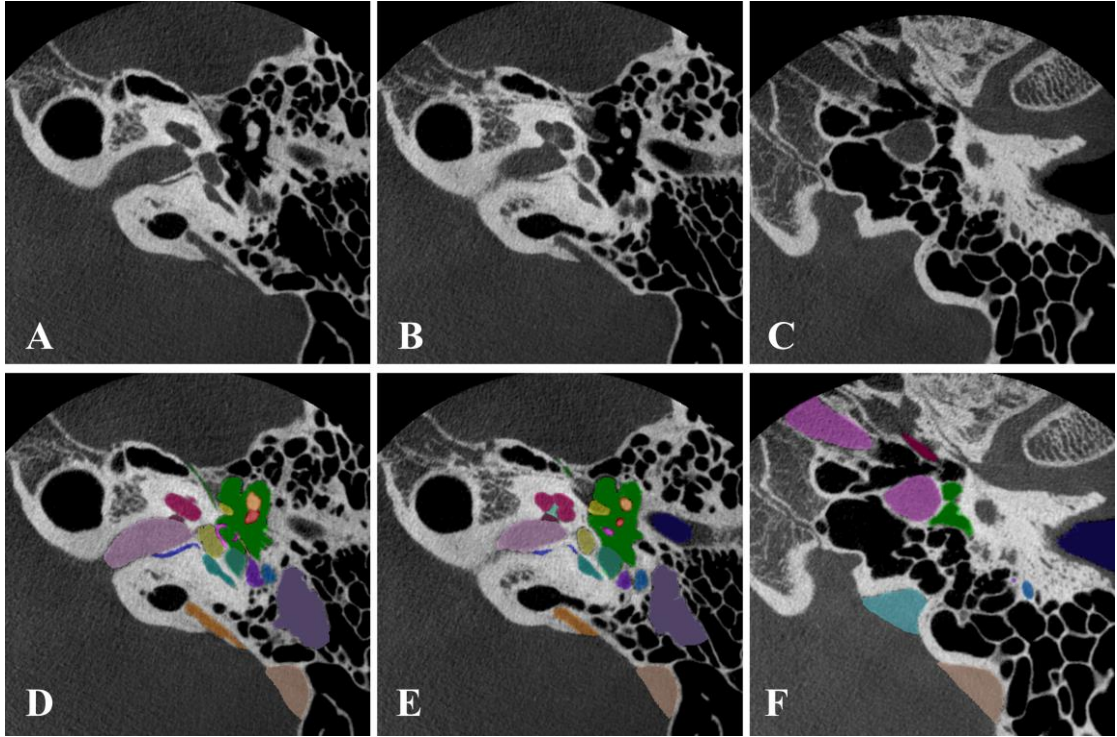

Figure 4. Image annotation in a healthy ear based on isotropic 0.1 mm U-HRCT (axial images), illustrating the anatomical structures of the external auditory canal, middle ear, and inner ear.

# Table 1. Standardized free-text CT report template for otologic imaging in the UltraEar Database

UltraEar Database: Ultra-High-Resolution Free-Text CT Report Template

| |
|---|
| Exam Name: UltraEar Database Ultra-High-Resolution CT <br><br> Exam Date: |
| Examination Technique: Ultra-high-resolution CT, axial sections, slice thickness/interval 0.1 mm |
| Imaging Findings:<br><br>Both external auditory canals are patent, with no abnormal changes in the bony walls. The tympanic membranes are not thickened. The mastoids are of the pneumatized type; good aeration of the temporal bone cells, mastoid antrum, and tympanic cavity is noted. No abnormal density is seen. The bony tegmen tympani and scutum are intact. The ossicles have normal morphology, density, and position. No abnormalities in the incudomalleolar or incudostapedial joints. All segments of the facial nerve canal are unremarkable. No abnormalities in the cochlea, vestibule, or semicircular canals. The round and oval windows are clearly visualized. Bone density around the cochlea is uniform without evidence of demineralization.<br>Both internal auditory canals are symmetric, without stenosis or dilatation. No enlargement of the vestibular aqueduct.<br>No high jugular bulb, no anteriorly displaced sigmoid sinus, and no low position of the middle cranial fossa on either side. |
| Impression:<br>No significant bony abnormality in the external, middle, or inner ear structures bilaterally. Correlate with clinical findings. |

## Declarations

### Ethics approval and consent to participate

All procedures performed in studies involving human participants were in accordance with the ethical standards of the local institutional review board of Beijing Friendship Hospital, Capital Medical University (approval No. 2018-P2-210-02, 2020-P2-061-02, 2022-P2-055-01 and 2024-P2-061-02) and with the 1964 Helsinki Declaration and its later amendments or comparable ethical standards.

### Consent for publication

Informed consent was obtained from all individual participants included in the study.

### Availability of data and materials

The datasets generated and/or analyzed during the current study are not publicly available due to patient privacy, ethical restrictions and controlled access policy, but are available from the corresponding author on reasonable request, as well as providing details of their research objectives, data security plan, and evidence of ethics approval.

### Competing interests

The authors declare that they have no competing interests.

### Funding

This study has received funding by National Natural Science Foundation of China (No. 82302282, Ruowei Tang; No. 61931013, Zhenchang Wang; No. 82171886, Pengfei Zhao), Beijing Friendship Hospital, Capital Medical University (No. YYZZ202336, Ruowei Tang), Beijing Natural Science Foundation (No. 7222301, Pengfei Zhao), Capital's Funds for Health Improvement and Research (No. 2022-1-1111, Zhenchang Wang), Beijing Scholar 2015 (No. [2015] 160, Zhenchang Wang), Beijing Key Clinical Discipline Funding (No. 2021−135, Zhenchang Wang).

### Authors' contributions

Conceptualization: RT, XL, ZW, ZZ, ZW, PZ

Study design: RT, XL, ZW, ZZ, HY, HD, NX, YC, SG, YL, ZW, PZ

Data acquisition and analysis: RT, ZZ, HY, HD, NX, YC, MZ

Interpretation of data: RT, SG, YL, MZ

Writing - original draft preparation: RT, XL, ZW

Approval of the submitted version: RT, XL, ZW, ZZ, HY, HD, NX, YC, SG, YL, ZW, PZ, MZ

Agreed both to be personally accountable for contributions and the accuracy or integrity of the work: RT, XL, MZ, ZW, ZZ, HY, HD, NX, YC, SG, YL, ZW, PZ

**Acknowledgements:** none